\documentclass{optica-article}
\journal{oe}

\articletype{Research Article}

\usepackage{lineno}

\newcommand{\p}{\partial}

\newcommand{\vg}{\textsl{g}}

\newcommand{\vta}{\vartheta}
\newcommand{\om}{\omega}

\newcommand{\ta}{\theta}

\newcommand{\wt}{\tilde}

\newcommand{\cH}{{\cal H}}

\newcommand{\cE}{{\cal E}}
\newcommand{\cW}{{\cal W}}

\newcommand{\be}{\begin{equation}}                                       
	\newcommand{\ee}{\end{equation}}
\newcommand{\ba}{\begin{eqnarray}}
	\newcommand{\ea}{\end{eqnarray}}
\newcommand{\bref}[1]{(\ref{#1})}

\newcommand{\bi}[1]{\bibitem{#1}}\newcommand{\lab}[1]{\label{#1}}

\newcommand{\bsub}{\begin{subequations}}                      
	\newcommand{\esub}{\end{subequations}}     

\begin{document}
\pagenumbering{roman}

\title{Frequency combs with multiple offsets in THz-rate microresonators}

\author{D.N. Puzyrev\authormark{1,2} and D.V. Skryabin\authormark{1,2,*}}

\address{\authormark{1}Department of Physics, University of Bath, Bath, BA2 7AY, United Kingdom\\
\authormark{2}Centre for Photonics and Photonic Materials, University of Bath, Bath, BA2 7AY, United Kingdom}

\email{\authormark{*}d.v.skryabin@bath.ac.uk} 



\begin{abstract}
Octave-wide frequency combs in microresonators are essential for self-referencing. However, it is difficult for the small-size and high-repetition-rate microresonators to achieve perfect soliton modelocking over the broad frequency range due to the detrimental impact of dispersion. Here we examine the stability of the soliton states consisting of one hundred modes in silicon-nitride microresonators with the one-THz free spectral range. We report the coexistence of fast and slow solitons in a narrow detuning range, which is surrounded on either side by the breather states. We decompose the breather combs into a sequence of sub-combs with different carrier–envelope offset frequencies. The large detuning breathers have a high frequency of oscillations associated with the perturbation extending across the whole microresonator. The small detuning breathers create oscillations localised on the soliton core and can undergo the period-doubling bifurcation, which   triggers a sequence of intense sub-combs.
\end{abstract}

\section{Introduction}

Broadband optical frequency combs find their applications in optical metrology, spectroscopy and optical information processing~\cite{rev1}. The high repetition rate, small footprint, and possibility of integrating the resonator and pump source on the same chip are the attractive features of the microresonator frequency comb technology~\cite{rev2}. 
Comb self-referencing requires achieving the octave-wide soliton spectral spans. Methods to generate the broad soliton combs rely on, e.g., reducing the microresonator radius~\cite{optica1,optica2}, higher order dispersion~\cite{mil,kip},   frequency doubling in a material with $\chi^{(2)}$ nonlinearity~\cite{hong} and
bi-chromatic pump~\cite{pascal,nist,mir}. 

\begin{figure*}[t]
	\centering{
		\includegraphics[width=1.\linewidth]{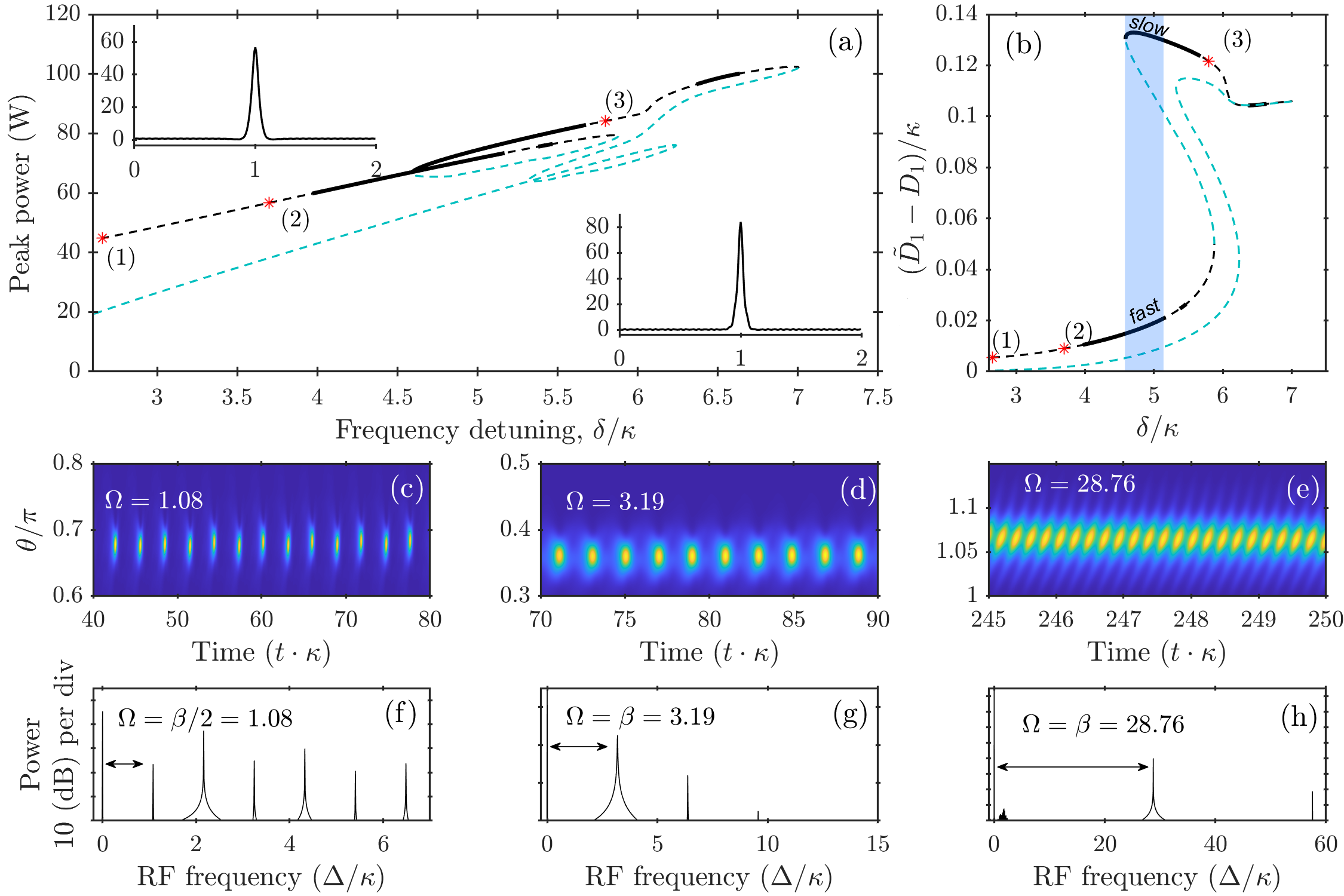}}
	\caption{(a) Soliton peak power as a function of the frequency detuning $\delta/\kappa$ for the fixed laser power $\cW = 29.5$mW. (b) Difference between the soliton repetition rate, $\wt D_1$, and the bare resonator rate, $D_1$. Full black lines in (a), (b) show stable solitons. Dashed-black and dashed-blue lines show unstable solitons. The insets in (a) show the stationary (unstable) soliton profiles vs $\ta/\pi$ at points (2) and (3). Panels (c)-(h) show the soliton breathers and their radio frequency (RF) spectra at points (1), (2), and (3). $2\pi/\Omega$ is the breather period. $\Omega$ is given in units of $\kappa$.}
	\lab{f1}
\end{figure*}	

In the microresonators, the comb lines start making their equidistant spectral steps from the pump laser frequency. However, this ideal grid does not always extend across the entire comb. The spectral structure of the octave combs should be carefully examined on a case-by-case basis because it can be composed of the overlapping combs having different repetition rates and different carrier–envelope offset frequencies~\cite{pascal,nist,offset,mir}. For the sake of brevity, we refer below to the carrier–envelope offset frequencies as the offset frequencies or offsets.
The  Significant changes in the dispersion and growth of specific modes triggering new four-wave mixing cascades can be the reasons behind this unwanted complexity. Therefore, developing more profound insights into the physical mechanisms behind the multiple offset frequencies and repetition rates is beneficial.

Our goal is to present a theoretical framework and associated numerical data using an example of the silicon nitride THz-rate resonator~\cite{optica1,optica2,pap1}. 
We report the coexisting stable soliton combs with different repetition rates, i.e., fast and slow solitons.
The soliton instability happens via the oscillatory scenario leading to the formation of soliton breathers~\cite{nb,sk,br1,br2,br3,brm,br4,br5}.  The range of the breather existence is broad and happens on either side from the pump detuning interval supporting stable solitons. For small detunings, the unstable waveforms are localised around the soliton core, while the large detuning instabilities are driven by the perturbations smeared along the resonator circumference. We demonstrate that the breather spectra are composed of several, sometimes very intense, overlapping combs having different offset frequencies and sharing the same repetition rates. Mathematics underpinning our formalism is described in Appendix, while the main text is focused on the data analysis.

\begin{figure}[t]
	\centering
	\includegraphics[width=0.8\linewidth]{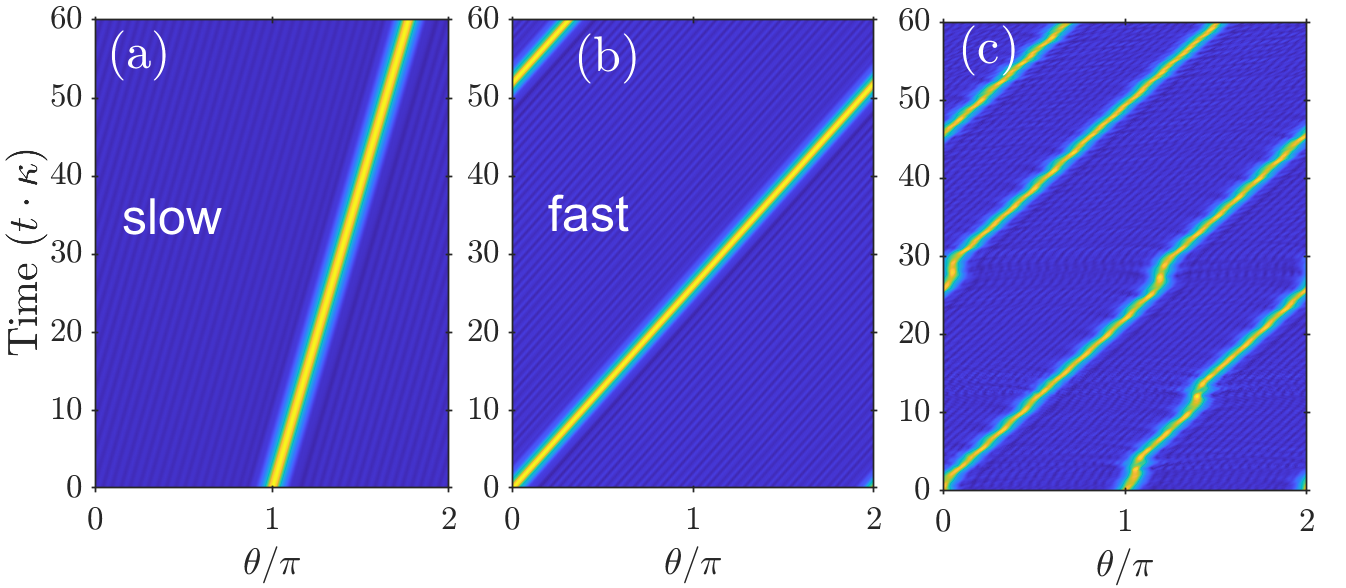}
	\caption{Space-time trajectories of the slow (a) and fast (b) solitons plotted in the reference frame rotating with the rate of the bare resonator, $D_1$. The laser power $\mathcal{W}=29.5$mW 
		and $\delta/\kappa=4.6$. (c) Random switching between the fast (longer time intervals) and slow (shorter intervals) for the two solitons simultaneously present in the resonator.}
	\lab{f2}
\end{figure}

\section{Fast and slow solitons}

We assume that the electric field, $\cE$, inside the resonator is
\be
\cE=e^{iM\vta-i\om_p t}\psi(t,\vta)+c.c.,
\lab{eff}
\ee
where $\om_p$ is the pump laser frequency, $t$ is time, $\vta\in[0,2\pi)$ is the angular coordinate 
in the laboratory frame, and $M$ is the mode number of 
the resonance nearest to $\om_p$. The envelope function $\psi$ is sought as the mode expansion in the rotating frame of reference, 
\be
\psi=\sum_{\mu=-N/2+1}^{N/2}\psi_\mu(t) e^{i\mu\ta},~\ta=\vta-\tilde D_1 t.
\lab{ef}
\ee
Here, $\psi_\mu$ are the complex modal amplitudes, 
$\mu$ are the relative mode numbers, and 
$\ta$ is the coordinate in the frame rotating with the 
repetition rate $\wt D_1$. 
We assume here that the bare resonator spectrum, $\om_\mu$, is approximated 
by $\omega_\mu=\omega_0+\mu D_1+\tfrac{1}{2!}\mu^2D_2+\tfrac{1}{3!}\mu^3D_3+\tfrac{1}{4!}\mu^4D_4
+\tfrac{1}{5!}\mu^5D_5+\tfrac{1}{6!}\mu^6D_6$, where
$\om_0/2\pi=193$THz. The 
dispersion parameters match the THz-rate Si$_3$N$_4$ resonator used in Ref.~\cite{pap1}, see Fig.~4 there: $D_1/2\pi=1$THz, $D_2/2\pi=20$MHz, $D_3/2\pi=0$Hz, $D_4/2\pi=-120$kHz, $D_5/2\pi=5.3$kHz, $D_6/2\pi=-150$Hz. The linewidth is $\kappa/2\pi=125$MHz.
$\psi_\mu(t)$ are driven by the well-established coupled-mode equations described in Appendix. 

Soliton frequency combs realise a particular case of Eq.~\bref{ef} such that
\be
\psi_\mu(t)=\hat\psi_\mu=\text{const}_\mu,~\psi=\sum_\mu\hat\psi_\mu e^{i\mu\ta}.
\lab{sol}\ee
The interplay of nonlinearity and higher-order dispersion makes the soliton repetition rate $\wt D_1$  to be different from the bare resonator rate value, $D_1$. The repetition rate difference, $\wt D_1- D_1$, depends on the resonator and pump parameters and has been determined numerically self-consistently with the modal amplitudes $\hat\psi_\mu$. 	
Numerical data assembled in Fig.~\ref{f1} trace the soliton peak power, $\max_\ta|\psi|^2$, and $\wt D_1- D_1$ as functions of the pump detuning, $\delta=\om_0-\om_p$, for the fixed input power $\cW=29.5$mW. 

The relatively large angular soliton width, $\approx 2\pi/10$, boosts the role played by the finite size effects. This makes the bifurcation structure of the THz-rate solitons to be more complex than the one in the 
case of low repetition rate resonators, see, e.g., Fig.~4 in Supplemental Materials of Ref.~\cite{herr}, 
which corresponds to the damped-driven Nonlinear Shr\"odinger equation (Lugiato-Lefever model) solved on the effectively infinite spatial interval~\cite{new,ba}. The complex behaviour of the soliton peak power in  Fig.~\ref{f1}(a) corresponds to the more transparent structure of the $\wt D_1- D_1$ vs $\delta$ plot in Fig.~\ref{f1}(b). Sufficiently rapid change of $\wt D_1- D_1$ around $\delta/\kappa=5$ separates the ranges of the slow and fast solitons. Numerical investigation of the power dependencies of the soliton families shows that the fast solitons emerge for powers above 
$\simeq 20$mW.

The stability of solitons relative to small perturbations is an important issue that we have analysed in detail using the formalism described in Appendix. The dashed and full lines in Figs.~\ref{f1}(a) and \ref{f1}(b) correspond to the unstable and stable solitons, respectively. It should be noted that the stability intervals of the fast and slow solitons overlap, see the shading in Fig.~\ref{f1}(b). The space-time trajectories of the slow and fast solitons in the reference frame rotating with the bare resonator rate $D_1$ are shown in Figs.~\ref{f2}(a) and \ref{f2}(b), respectively. If the model is initialised simultaneously with the fast and slow solitons, e.g., located on the opposing sides of the resonator, see Fig.~\ref{f2}(c), then 
we have observed the convergence of the slow soliton to the fast one. This process could not, however, fully stabilise itself, and the pair was making random switches to the slow state and back to fast. 

\section{Soliton breathers and combs with multiple offsets}

The unstable soliton branches marked with the blue dashed lines represent the so-called low-branch solitons. If the model is initialised with these solitons, it collapses to the single mode state, $\mu=0$. The solitons shown by the dashed black lines are unstable to the oscillatory (Hopf) instability. The waveforms driving the Hopf instability are either localised on the soliton core (small $\delta$'s) or extend along the entire resonator circumference (large $\delta$'s), see Appendix. Typical operation regimes emerging from the initial conditions on the black dashed lines correspond to the soliton breathers; see Figs.~\ref{f1}~(c)-(e). 

The breather dynamics computed at  points (1),(2) and (3), see  Figs.~\ref{f1}(a),(b),  
is shown in Figs.~\ref{f1}(c),(d) and (e). The reference frame is chosen to rotate with the breather rate,
which is generally different from the rate of the unstable stationary solitons taken for the same parameters.
Figs.~\ref{f1}(f)-(h) show the RF (power) spectra computed in the rotating frame,
$\int_0^\tau  \sum_\mu  | \psi_\mu |^2 e^{-i\Delta t}dt/\tau $. 
Points (2) and (3) are nearest to the stable solitons, and the  RF spectra are near harmonic. 
Here, the breather frequency $\Omega$ practically coincides with the 
imaginary part $\beta$ of the eigenvalues driving instabilities of the stationary solitons, 
$e^{\lambda t-i\beta t}$, see Appendix for details. 
Point (1) is the furthest from the Hopf threshold
and is beyond the period doubling threshold also. Here, the breather frequency becomes  
$\Omega=\beta/2$. The primary oscillations are still happening with $\beta$, 
but the small up and down shifts of the neighbouring intense spots,
see Fig.~\ref{f1}(c), double the period and half the breathing frequency.

\begin{figure}[t]
	\centering
	\includegraphics[width=0.8\textwidth]{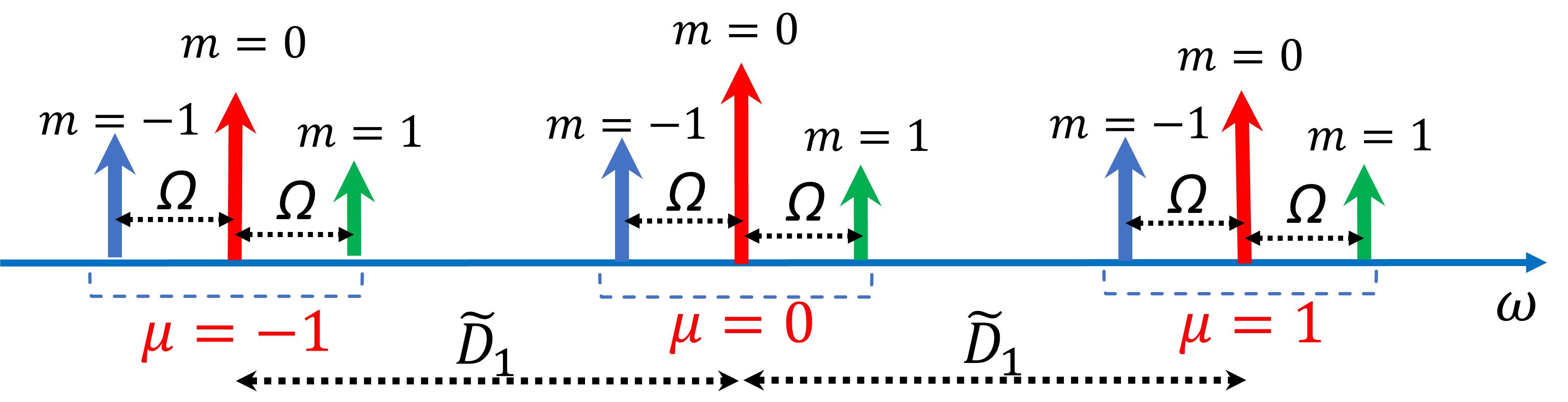}
	\caption{Illustration of the breather spectrum. The breathing period is $2\pi/\Omega$. $m=0$ is the primary comb and $m=\pm 1$ are the sub-combs. The repetition rate, $\wt D_1$, is the same for the primary and all sub-combs, while the offset frequencies differ by $m\Omega$.  }
	\lab{f3}
\end{figure}	

\begin{figure}[t]
	\centering
	\includegraphics[width=0.8\linewidth]{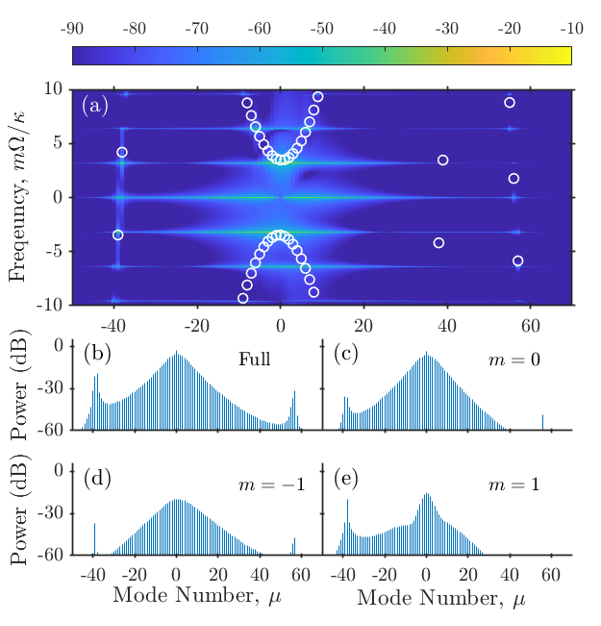}
	\caption{ Panel (a) shows the two-dimensional map representing the power spectrum of the soliton breather, $|\hat \psi_{\mu,m}|^2$, computed at the point (2) in Fig.~\ref{f1}, see Eq.~\bref{br}. The horizontal lines are separated by the breather frequency $\Omega/\kappa=3.19$. The small panels below show the full (b), the primary (c), $|\hat \psi_{\mu,0}|^2$, and the higher order, $|\hat \psi_{\mu,|m|>0}|^2$, sub-combs; see Eq.~\bref{subc}. White circles in (a) show the spectrum of excitations around the single mode solution, $\mu=0$, see Eq.\bref{bog}. The colour map interpolation obscures the discreteness of the spectrum in (a).}
	\lab{f4}
\end{figure}

\begin{figure}[t]
	\centering
	\includegraphics[width=0.8\linewidth]{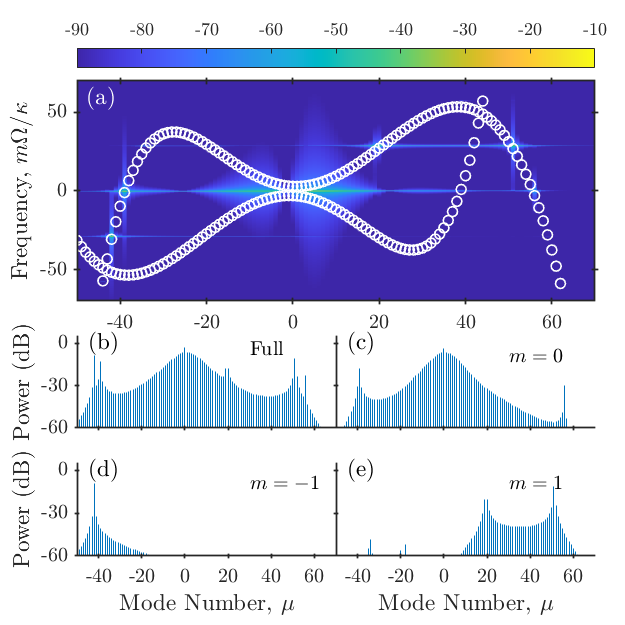}
	\caption{ Panel (a) shows the two-dimensional map representing the power spectrum of the soliton breather, $|\hat \psi_{\mu,m}|^2$, computed at the point (3) in Fig.~\ref{f1}, see Eq.~\bref{br}. The horizontal lines are separated by the breather frequency $\Omega/\kappa=28.76$. The small panels below show the full (b), the primary (c), $|\hat \psi_{\mu,0}|^2$, and the higher order, $|\hat \psi_{\mu,|m|>0}|^2$, sub-combs; see Eq.~\bref{subc}. White circles in (a) show the spectrum of excitations around the single mode solution, $\mu=0$, see Eq.\bref{bog}. The colour map interpolation obscures the discreteness of the spectrum in (a). Note, the difference of the y-axis scale in (a) here and in Fig.~\ref{f4}. }
	\lab{f5}
\end{figure}

Since the breather combs have periodic modal amplitudes, 
\be
\psi_\mu(t)=\hat\psi_\mu(t)=\hat\psi_\mu(t+2\pi/\Omega),
\ee
the time-domain Fourier expansion of $\hat\psi_\mu(t)$ yields 
\be
\psi=
\sum_\mu\left[\sum_{m}\hat\psi_{\mu,m}e^{-im\Omega t}\right] e^{i\mu\ta},
\lab{br}
\ee
where $\hat\psi_{\mu,m}$ are the time-independent amplitudes of the breather spectrum. 
Using Eq.~\bref{eff} and Eq.~\bref{br} and rearranging we find that the electric field of a breather is
\be
\cE
=\sum_m\left[e^{iM\vta-i(\om_p+m\Omega) t}\sum_\mu\hat\psi_{\mu,m} 
e^{i\mu\ta}\right]+c.c..\lab{br1}
\ee
Thus,  the breather
is a superposition of the primary, $m=0$, and higher order, $m\ne 0$, sub-combs 
which all share the same repetition rate,
$\wt D_1$, but have different central frequencies $\omega_p+m\Omega$, and, therefore, different offsets from zero. The multiple spectral grids of the breather combs are represented by
\be
\om_{\mu,m}=\om_p+m\Omega+\mu \wt D_1,~m=0,\pm 1,\pm 2,\dots,
\lab{subc}
\ee with $|\hat\psi_{\mu,m}|^2$ being the sideband powers in the $m$'s subcomb, 
see Fig.~\ref{f3} for an illustration.

The colour map in Fig.~\ref{f4}(a) shows the dB levels of the $|\hat\psi_{\mu,m}|^2$ 
spectrum for the breather found at the point (2), i.e., on the left from the soliton stability interval in Fig.~\ref{f1}(a).
The discreteness of the horizontal lines in Fig.~\ref{f4}(a) is smoothed over 
by the colour map interpolation. The vertical separation of these lines equals $\Omega$.
The upward, $\beta_\mu^+$, and downward, $\beta_\mu^-$, 
parabolas, and the sparsely separated circles for larger $\mu$
show the spectrum of the stable small amplitude excitations, $e^{(\lambda_\mu^\pm-i\beta_\mu^\pm)t}$, 
$\lambda_\mu<0$, existing on top of the single mode, $\mu=0$, state,
\be
\begin{split}
	\beta_\mu^\pm=&\tfrac{1}{2}\left(\om_\mu-\om_{-\mu}-2\mu\wt D_1\right)\\
	\pm&\sqrt{3\left(\vg-\tfrac{1}{2}(\om_\mu+\om_{-\mu}-2\om_p)\right)\left(\vg-\tfrac{1}{6}(\om_\mu+\om_{-\mu}-2\om_p)\right)}.
\end{split}\lab{bog}
\ee
Here, $\vg\sim |\psi_0|^2$ is the nonlinear shift of the resonance; see Appendix for further details.

Data in Fig.~\ref{f4}(a) bring us to an immediate  conclusion 
that, in this instance, the breather frequency $\Omega$ equals 
$\beta_0^{+}=\sqrt{3(g-\delta)(g-\delta/3)}$. Hence, the primary physical reason 
for breathing comes from the resonance excitation of the group of modes around $\mu=0$.
This is further corroborated by the spectral shapes of the $m=\pm 1$ 
sub-combs, where the central group of modes is either dominant,  
see Fig.~\ref{f4}(d), or more pronounced than the $\mu=-40$ group, see Fig.~\ref{f4}(e).

Now, we are moving onto describing the sub-combs driving the breather dynamics 
for large detunings, i.e., on the right from the soliton stability interval in Fig.~\ref{f1}(a).
The colour map of $|\hat\psi_{\mu,m}|^2$ in Fig.~\ref{f5}(a) shows that
the breather frequency $\Omega$ is now much larger and is given
by $\beta_{52}^{+}\approx\beta_{20}^{+}\approx\beta_{-42}^{+}$. 
The shape of the $m=\pm 1$ sub-combs in Figs.~\ref{f5}(d), (e) is unambiguous 
that spectral groups around $\mu=-42,20$, and $52$ are now dominant in the breather structure,
and the spectrum around $\mu=0$ is practically absent.

The higher-order sub-combs with $|m|\ge 2$ have very small power 
in the examples presented in Figs.~\ref{f4}, \ref{f5}. This is because points (2) and (3) 
are located not deep enough into the instability range, see Fig.~\ref{f1}(a). However, point (1) corresponds to well-developed instability, 
which has already crossed the period doubling threshold. 
The higher-order sub-combs become well developed here, see 
the  $m=0,\pm 1,\pm 2,\pm 3$, and $\pm 4$ cases in Fig.~\ref{f6}.
Predictably, the even order sub-combs are more powerful
because they correspond to a subsequence of the stronger peaks in the RF spectrum in Fig.~\ref{f1}(f). Figs.~\ref{f4}-\ref{f6} show the mode-number spectra, while the optical frequencies, as measured on a spectrum analyser, must be found using Eq.~\bref{subc}. Fig.~\ref{fnew}, however, shows the selected spectral lines and a whole of the numerically computed frequency spectrum corresponding to the mode-number data in Fig.~\ref{f5}. Zooming on the individual lines reconstructs the schematic illustration in Fig.~\ref{f3}. The predicted spectral features are amenable to measurements on the high resolution spectrum analysers. Possible oscillatory instabilities due to interaction of different mode families may yield higher values of $\Omega$.

\begin{figure}[t]
	\centering
	\includegraphics[width=0.8\linewidth]{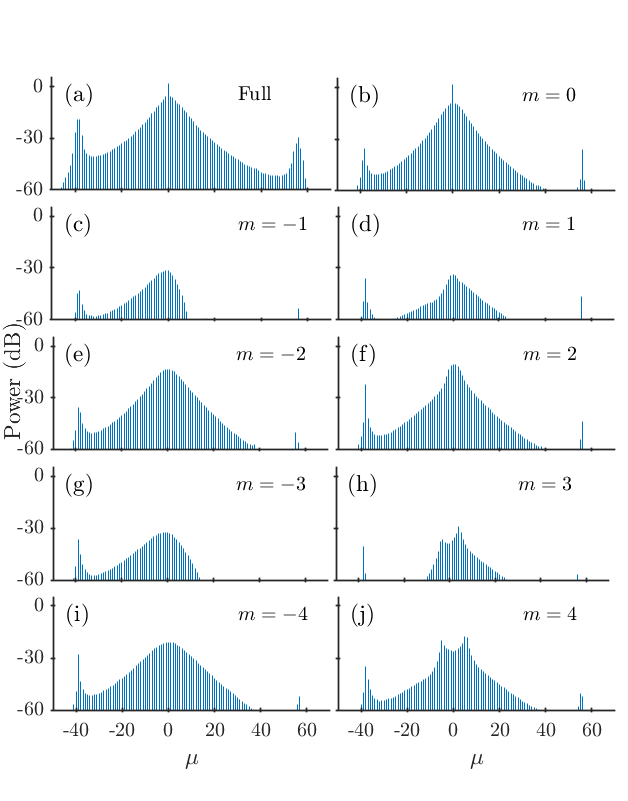}
	\caption{  Full (a),  primary (b) and  higher-order sub-combs computed away from the Hopf instability boundary at point (1) in Fig.~\ref{f1}. }
	\lab{f6}
\end{figure}

\begin{figure}[t]
	\centering
	\includegraphics[width=0.8\linewidth]{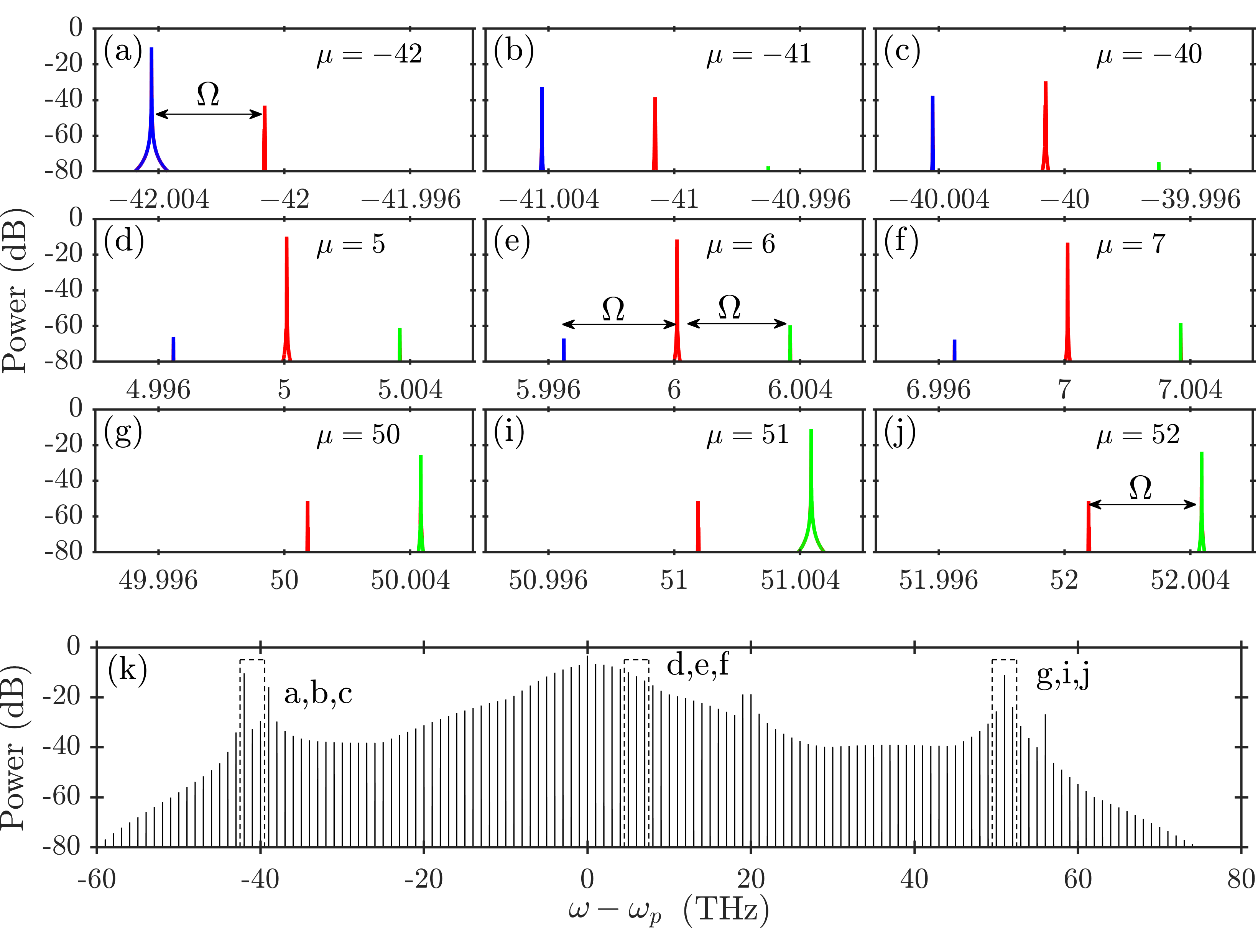}
	\caption{Selected spectral lines (a-j) and a whole of the numerically computed frequency spectrum (k) corresponding to the mode-number data in Fig.~\ref{f5}. Zooming on the individual spectral lines  reconstructs the schematic illustration in Fig.~\ref{f3}.  The red lines in (a-j) correspond to the primary $m=0$ comb. The green and blue lines are the $m=1$ and $m=-1$ sub-combs: $\Omega/2\pi=3.6$GHz. }
	\lab{fnew}
\end{figure}

\section{Summary}
Soliton breather states in the damped and driven nonlinear Schr\"odinger (Lugiato-Lefever) equation have been known for a long time~\cite{nb,sk}. The recent resurgence in studies of their properties has happened primarily through the context of the frequency comb generation in 
microresonators~\cite{pascal,nist,offset,mir,br1,br2,br3,brm,br4,br5}. What was particularly interesting for us to find and report here is the significant power and bandwidth of the breather sub-combs with different offset frequencies. The modal composition of these sub-combs varies between the ones spectrally localised around the pump and others detuned away. These detuned sub-combs are associated with relatively high-frequency breathers. All this complexity happens through the interplay of the large second and higher order dispersions and a relatively small number of modes covering the significant bandwidth, which are all features of the small-size and high-repetition-rate microresonators. Stable solitons can also be found under this extreme spectral broadening. The bifurcation structure of the THz-rate solitons and breathers differs from the one found in the idealised Lugiato-Lefever equation solved on the infinite spatial domain. In addition to the breathers found at the small and large detunings, we have reported the coexisting stable solitons having two different repetition rates, i.e., slow and fast solitons.

\begin{backmatter}
	
	\bmsection{Funding} Engineering and Physical Sciences Research Council (2119373); Horizon 2020 Framework Programme (812818).
	
	\bmsection{Acknowledgments}  Authors thank S.B. Papp for sharing and discussing experimental data.
	
	\bmsection{Disclosures} The authors declare no conflicts of interest.
	
	\bmsection{Data Availability Statement}   Data underlying the results presented in this paper are not publicly available but may be obtained from the  authors upon reasonable request.

\end{backmatter}

\section*{Appendix}

The starting point of the derivation of our model is the Maxwell equations in the ring-resonator geometry and with Kerr nonlinearity~\cite{herr,chembo,osacont}.
The modal amplitudes $\psi_\mu$ and the field envelope $\psi$, see Eq.~\bref{ef},
comply with the coupled-mode 
equations that can be cast to the following form~\cite{osacont},
\be
i\p_t\psi_\mu=\Big(\om_\mu-\om_p-\mu \wt D_1\Big)\psi_\mu
-\frac{i\kappa}{2}\left(\psi_\mu-\hat\delta_{0,\mu}\cH\right)
-\gamma \int_0^{2\pi}\psi |\psi|^2 e^{-i\mu\ta}\frac{d\ta}{2\pi}.\lab{fun}
\ee
Eq.~\bref{fun} contains several parameters in addition to the ones defined in the main text.
$\cH$ is the pump parameter linked to the laser power, $\cW$, as $\cH^2=\eta\cW D_1/2\pi \kappa$, where  
$\eta$ is the dimensionless coupling parameter. $\hat\delta_{\mu',\mu}=1$ for $\mu=\mu'$ and $\hat\delta_{\mu',\mu}=0$ otherwise, and $\gamma$ is the nonlinear parameter, $\gamma/2\pi=20$MHz/W.
The $\psi_\mu$ representation of the nonlinear term yields~\cite{osacont},
\be
i\p_t\psi_\mu=\Big(\om_\mu-\om_p-\mu \wt D_1\Big)\psi_\mu
-\frac{i\kappa}{2}\left(\psi_\mu-\hat\delta_{0,\mu}\cH\right)
-\gamma \sum_{\mu_1\mu_2\mu_3} 
\hat\delta_{\mu_1+\mu_2-\mu_3,\mu}
\psi_{\mu_1}\psi_{\mu_2}\psi_{\mu_3}^*.\lab{fu1}
\ee
The stationary soliton solutions, $\psi_\mu=\hat\psi_\mu$, see Eq.~\bref{sol}, have been found using the Newton method.

To differentiate between stable and unstable 
solitons and identify the oscillatory (Hopf) instabilities 
we have analysed the linear stability of the soliton family. 
We perturbed the time-independent 
soliton amplitudes, $\hat\psi_\mu$, with small perturbations, 
$\epsilon_\mu(t)=x_\mu(t)+y_\mu^*(t)$, 
\be
\psi_\mu(t)  = \hat\psi _\mu  + x_\mu(t) +y^*_\mu( t),
\ee
and then linearised Eq.~\bref{fu1}, 
\be
\begin{split}
\lab{s1}
i(\p_t+\tfrac{1}{2}\kappa)x_\mu&=
\Big(\om_\mu-\om_p-\mu \wt D_1\Big)x_\mu\\
&-\gamma \sum_{\mu_1\mu_2\mu_3} 
\hat\delta_{\mu_1+\mu_2-\mu_3,\mu}
\left(
\psi_{\mu_2}\psi_{\mu_3}^*x_{\mu_1}+
\psi_{\mu_1}\psi_{\mu_3}^*x_{\mu_2}+
\psi_{\mu_1}\psi_{\mu_2}y_{\mu_3}
\right),
\\
i(\p_t+\tfrac{1}{2}\kappa)y_\mu&=
-\Big(\om_\mu-\om_p-\mu \wt D_1\Big)y_\mu\\
&+\gamma \sum_{\mu_1\mu_2\mu_3}
\hat\delta_{\mu_1+\mu_2-\mu_3,\mu}
\left(
\psi_{\mu_1}^*\psi_{\mu_2}^*x_{\mu_3}+
\psi_{\mu_1}^*\psi_{\mu_3}y_{\mu_2}+
\psi_{\mu_2}^*\psi_{\mu_3}y_{\mu_1}
\right).
\end{split}
\ee
By substituting $\left\{x_\mu(t),y_\mu(t)\right\}=
\left\{X_\mu,Y_\mu\right\}e^{\lambda t-i\beta t}$, we have reduced Eq.~\bref{s1} to the algebraic $2N\times 2N$ eigenvalue problem, which was solved numerically.
 The soliton is stable if 
all $\lambda<0$ and becomes Hopf unstable when a pair of identical $\lambda$'s  
with $\pm\beta\ne 0$ crosses zero. 
Near the Hopf threshold, $\beta$ 
closely matches the breather frequency $\Omega$, see Figs.~\ref{f1}(g), (h). Components of the eigenvectors in physical space are expressed as
$X(\ta)=\sum_\mu X_\mu e^{i\mu\ta}$
and $Y(\ta)=\sum_\mu Y_\mu e^{i\mu\ta}$.
\begin{figure}[t]
	\centering
	\includegraphics[width=1.\linewidth]{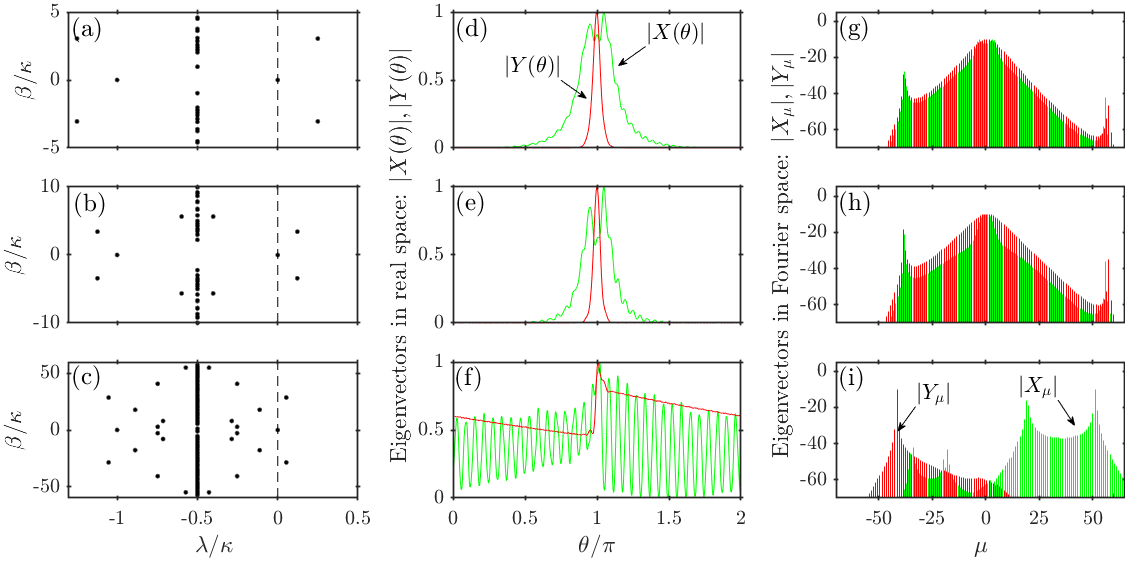}
	\caption{Eigenvalues and eigenvectors of the perturbations of the solitons unstable relative to the breather excitations. The first, second and third rows correspond to the points (1), (2) and (3) in Fig.~\ref{f1}, respectively. Each panel in the first column has a distinct pair of unstable eigenvalues, $\lambda>0$. The second and third columns show the corresponding eigenvectors in the physical and Fourier spaces, respectively. }
	\lab{f7}
\end{figure}

Numerically computed eigenvalues of the  
unstable solitons at points (1), (2) and (3) in Fig.~\ref{f1}(a) are shown in Figs.~\ref{f7}(a)-(c). As one can see, all three cases correspond to the Hopf instability with the pairs of complex conjugate eigenvalues. For points (1) and (2), i.e., left from the soliton stability interval, the eigenvectors are localised around the soliton core and are largely shaped by the group of modes around $\mu=0$, see Fig.~\ref{f7}. 
For point (3), i.e., right from the soliton stability interval, the eigenvectors are spread across the entire resonator circumference and spectrally centred around $\mu=-42,20$, and $52$. The eigenvector analysis supports the conclusions from decomposing the dynamical data into sub-combs described in the main text.

Frequencies of the excitations supported by the single mode, $\psi_{\mu\ne 0}=0$, state, see Eq.~\bref{bog}, are found from the simple two-by-two eigenvalue problem, 
\begin{equation}
	i\left(\lambda_\mu^\pm-i\beta_\mu^\pm+\tfrac{1}{2}\kappa\right)\begin{bmatrix} x_\mu\\ y_{-\mu}
		\end{bmatrix}=\begin{bmatrix}\displaystyle
		(\om_{\mu}-\om_p-\mu\wt  D_1)-2\vg
		&\displaystyle -\vg e^{i\phi}\\
		\displaystyle\vg  e^{-i\phi}& 
		\displaystyle -(\om_{-\mu}-\om_p+\mu\wt  D_1)+2\vg
	\end{bmatrix}
\begin{bmatrix} x_\mu\\ y_{-\mu}
\end{bmatrix}.
	\lab{hh}
\end{equation}
Here, $\vg=\gamma |\psi_0|^2$ and $\vg e^{i\phi}=\gamma\psi_0^2$. In this case, each eigenvalue is attributed to a particular pair of sidebands, $\pm\mu$. The '$\pm$' superscripts in the left side of Eq.~\bref{hh} are used to distinguish between the two eigenvalues of the matrix, see Eq.~\bref{bog}.


\end{document}